\documentclass[a4paper]{article}
\usepackage{ISCSLP2026}
\usepackage{url}
\usepackage{enumitem}
\usepackage{etoolbox}

\title{E3-System: Expressive EmoData, Explicitly Controllable, Easy-to-Use TTS}
\usepackage{pifont}

\name{
Yiqiao Huang$^{1,*}$,
Xiaolong Lin$^{2,*,\dagger}$,
Zhenpeng Zhan$^{2}$,
Zhizheng Wu$^{3}$
}

\address{
$^{1}$Harvard University, Cambridge, MA, USA\\
$^{2}$Baidu Inc., Shenzhen, China\\
$^{3}$The Chinese University of Hong Kong, Shenzhen, China
}

\email{
linxiaolong01@baidu.com
}

\usepackage{comment}
\usepackage{tabularx}
\usepackage{float}

\begin{document}
%

\maketitle
\begingroup
\renewcommand{\thefootnote}{\fnsymbol{footnote}}
\footnotetext[1]{\tiny Equal contribution.\hspace{1em}$\dagger$ Corresponding author.}
\endgroup
\vspace*{-2.0em}
\begin{abstract}

Recent zero-shot Text-to-Speech (TTS) systems can clone unseen speakers, yet controllable expressiveness remains limited by a gap between semantic intent and acoustic realization. Text descriptions offer a natural control interface but provide weak acoustic grounding, whereas speech style prompts provide concrete prosody but entangle style with the prompt speaker's timbre.
To bridge this gap, we propose E3-System, a unified intent-to-speech framework for expressive zero-shot TTS that couples three requirements: \textbf{E}xpressive style modeling, \textbf{E}xplicit timbre-style control, and \textbf{E}asy-to-use natural-language interaction. Given a target speaker and either a natural-language style description or a speech style prompt, E3-System generates expressive speech while preserving the target timbre. The key idea is to close the full control loop: building a large expressive style space, grounding user intent into usable acoustic style references, and synthesizing speech with disentangled timbre and style. To realize this loop, EmoData provides a thousand-hour multilingual emotional speech dataset, Style-RAG grounds text descriptions into high-quality acoustic style prompts from these data, and Dual-TTS transfers the grounded style to arbitrary target speakers through VC-augmented timbre-style disentanglement. Experiments show that E3-System achieves strong speech quality, speaker preservation, and explicit expressive controllability.
\end{abstract}

\section{Introduction}
\label{sec:intro}

Recent advances in zero-shot Text-to-Speech (TTS) have not only achieved near-human naturalness but also explored controllability for timbre and style. \cite{du2024cosyvoicescalablemultilingualzeroshot,du2024cosyvoice2scalablestreaming, zhang2025vevocontrollablezeroshotvoice,zhou2025indextts2breakthroughemotionallyexpressive,li2026discospeechcontrollablezeroshotspeech} {However, they typically face trade-offs between synthesis expressiveness and control flexibility, or rely on speech references, lacking the ease of explicit natural language control.} Thus, a practical expressive TTS system must bridge the gap between high-level user intent and faithful acoustic realization, while preserving speaker identity and supporting arbitrary style transfer.

Achieving this goal requires a complete control loop: a rich expressive style space, reliable grounding from user intent to acoustic style, and synthesis that disentangles style from speaker identity.
A key bottleneck lies in the expressive training data. Current large-scale TTS corpora are primarily derived from audiobooks or podcasts \cite{langman2025hifitts2largescalehighbandwidth,zen2019librittscorpusderivedlibrispeech,ma2024wenetspeech4tts12800hourmandarintts,he2024emiliaextensivemultilingualdiverse}. These datasets are clean and extensive, but mostly neutral in style. In contrast, studio emotional speech datasets \cite{DBLP:journals/lre/BussoBLKMKCLN08,adigwe2018emotional,Xin_2024} contain strong emotional expression but are much smaller in scale and may contain semantic-emotion mismatches (e.g., sad text acted with angry prosody). More recent resources \cite{abilbekov2024kazemottsdatasetkazakhemotional,zhou2022emotionalvoiceconversiontheory} partially address this gap, yet remain limited in scale and multilingual coverage.

Another challenge is controllable modeling of timbre and style.
Recent state-of-the-art zero-shot TTS systems \cite{wang2023neuralcodeclanguagemodels,ju2024naturalspeech3zeroshotspeech,anastassiou2024seedttsfamilyhighqualityversatile,guo2025fireredttsfoundationtexttospeechframework, wang2024maskgctzeroshottexttospeechmasked, wang2025sparkttsefficientllmbasedtexttospeech} have achieved remarkable fidelity in reconstructing both timbre and style from a single reference speech. However, this capability relies on the inherent coupling of speaker identity and prosody within the prompt. Consequently, they lack the flexibility to arbitrarily combine different timbres and styles. While disentanglement-based methods \cite{zhang2025vevocontrollablezeroshotvoice,zhang2023iemottsrobustcrossspeakeremotion,huang2024debattszeroshotdebatingtexttospeech} attempt to decouple these attributes, they often suffer from trade-offs between controllability and quality, manifesting as timbre, style, or content leakage. Ideally, a system should offer explicit disentanglement without compromising the generation quality seen in coupled SOTA baselines. E3-System addresses this requirement through Dual-TTS, which enables explicit timbre-style control while maintaining strong zero-shot performance.

Instruction paradigms also pose challenges. Text-instruct methods \cite{guo2022promptttscontrollabletexttospeechtext,yang2023instructttsmodellingexpressivetts, leng2023prompttts2describinggenerating} are easy to use, as users can simply describe desired timbre or style in natural language. {However, pure text instructions inherently lack the acoustic reference required for zero-shot voice cloning. Moreover, they cannot convey fine-grained acoustic details such as emotional intensity or rhythm, leading to high generation variance and instability.} The same textual description often yields inconsistent prosodic outcomes across attempts. Wav-instruct methods can preserve these acoustic details, but they rely on users to manually select appropriate reference speech. For non-expert users, this process is costly and error prone. Previous work has shown that inappropriate reference can severely degrade synthesis quality \cite{luo2025autostyle}. Recent studies on multimodal prompting and retrieval-augmented approaches \cite{luo2025autostyle,guan2024mmttsmultimodalpromptbased} partially address this problem, but they still struggle to balance usability and synthesis quality.

Therefore, we propose E3-System, a unified intent-to-speech framework that closes the full control loop from expressive style-space construction, to intent-based acoustic grounding, to timbre-style disentangled synthesis.
\begin{enumerate}[label=(\arabic*), leftmargin=*, itemsep=0pt, parsep=0pt, topsep=0pt]
    \item \textbf{Expressive style dataset}: We introduce EmoData, to our knowledge the first thousand-hour multilingual dataset of highly expressive emotional speech. Built via a scalable automated pipeline, EmoData bridges the gap between the scale and diversity of in-the-wild data, and the quality and expressiveness of studio recordings.
    \item \textbf{Explicit timbre-style synthesis}: We propose Dual-TTS, a dual-prompt architecture trained with a novel VC-augmented disentanglement strategy. This approach effectively disentangles timbre from style, enabling ``timbre $\times$ style'' zero-shot synthesis while minimizing the leakage issues inherent in previous disentanglement methods.
    \item \textbf{Easy-to-use Style-RAG}:
    We design a retrieval-augmented module that maps natural language intents to high-quality style references from the expressive style space. This aligns abstract user intent with fine-grained acoustic control.
    
\end{enumerate}

In summary, E3-System unifies expressive style-space construction, intent-based acoustic grounding, and timbre-style disentangled synthesis into a single intent-to-speech framework for expressive zero-shot TTS.

\section{System Architecture}
\label{sec:format}

We present E3-System, a dual-reference intent-to-speech framework for expressive zero-shot TTS. It turns user style intent into timbre-preserving expressive speech through a unified pipeline: EmoData provides a large expressive style space, Style-RAG retrieves acoustic style references from this space, and Dual-TTS transfers the retrieved style to arbitrary target speakers with disentangled timbre and style control.

\begin{figure*}
    \centering
    \includegraphics[width=0.9\linewidth]{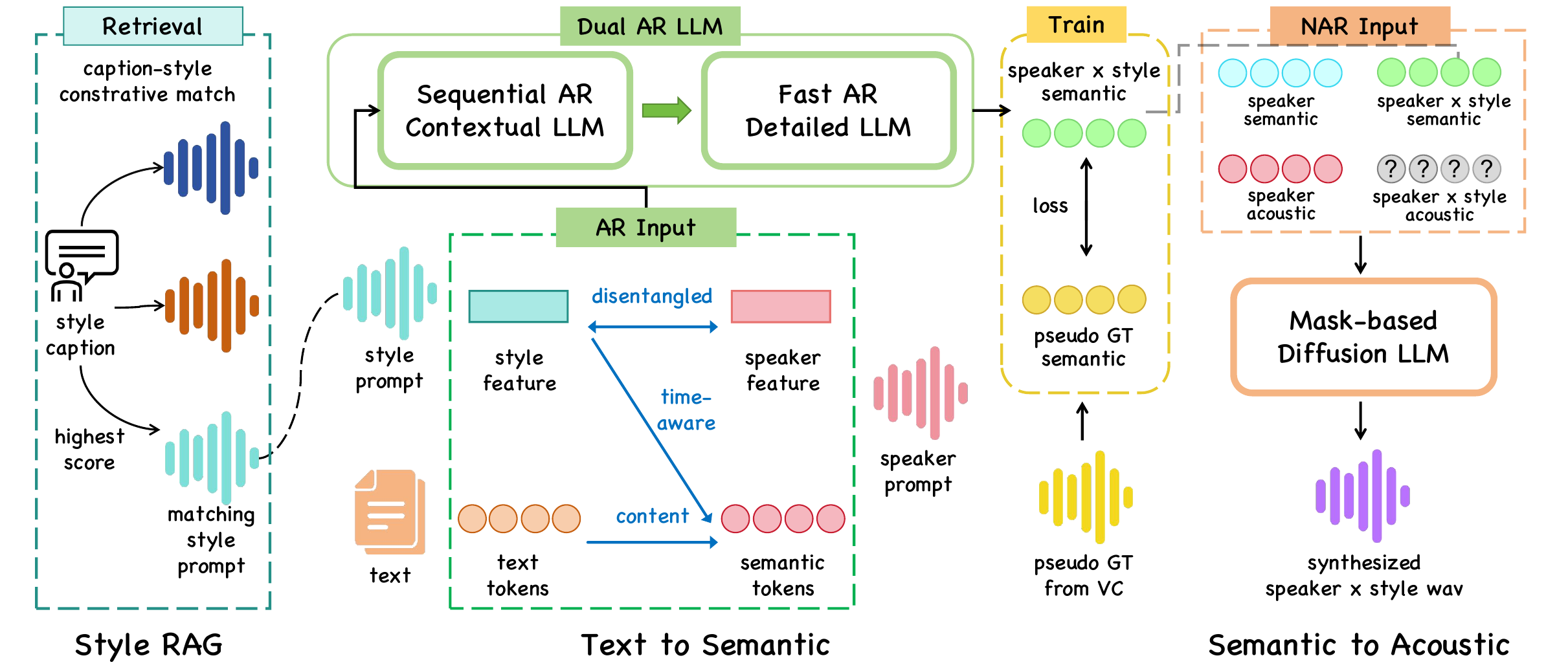}
    
    \caption{An illustration of the proposed Dual-TTS.}
    \label{fig:model}
\vspace{-5mm}
\end{figure*}

\subsection{Dual-TTS: Timbre-Preserving Style Transfer}
As the synthesis engine of E3-System, Dual-TTS takes two references as input: a speaker prompt for target timbre and a style prompt for expressive control. As shown in Fig~\ref{fig:model}, it contains two stages: text-to-semantic generation and semantic-to-acoustic generation.
\subsubsection{Text-to-Semantic Generation}
We construct conditional inputs for the generation of semantic tokens using input text, a style prompt, and a speaker prompt. The speaker prompt is encoded as a global embedding to control timbre. Meanwhile, the style prompt is encoded into a style vector and injected into reference semantic tokens to guide the speaking style. Unlike GRL-based disentanglement methods \cite{huang2023w2vc}, we adopt a data-driven decoupling strategy described below. To advance efficiency and consistency, we adopt a Dual AutoRegressive (Dual-AR) architecture specifically for the Text-to-Semantic stage, comprising a Contextual AR and a Fast AR. The Contextual-AR progressively captures long-range dependencies to maintain semantic consistency. The Fast-AR performs lightweight refinement, thereby accelerating inference while retaining local detail. Dual-TTS uses a 12-layer Contextual-AR, a 12-layer Fast-AR with 2$\times$ downsampling, all with 4096 hidden dimensions and 16 attention heads.

\subsubsection{Semantic-to-Acoustic Generation}

The model maps semantic tokens to acoustic tokens, which are then converted into waveforms by a decoder. We adopt a mask-based diffusion framework following SoundStorm\cite{borsos2023soundstormefficientparallelaudio}. Specifically, the inputs consist of the semantic and acoustic token extracted from the speaker prompt, and the speaker style semantic tokens predicted from the first stage. During training, a subset layers of acoustic tokens is randomly masked, and the model predicts masked parts conditioned on the semantic and unmasked acoustic tokens. This enables the model to learn both the mapping from semantics to acoustics and the recovery of fine-grained details. Our training schedule yields coarse-to-fine acoustic generation: the model first predicts the coarse acoustic pattern and then progressively refines it with local details to form a complete acoustic sequence. All models are pretrained on compliant, enterprise-licensed speech data, then fine-tuned on EmoData as described.

\subsubsection{Disentangle Training Paradigm}

\begin{figure}[t]
    \centering
    \includegraphics[width=0.88\linewidth]{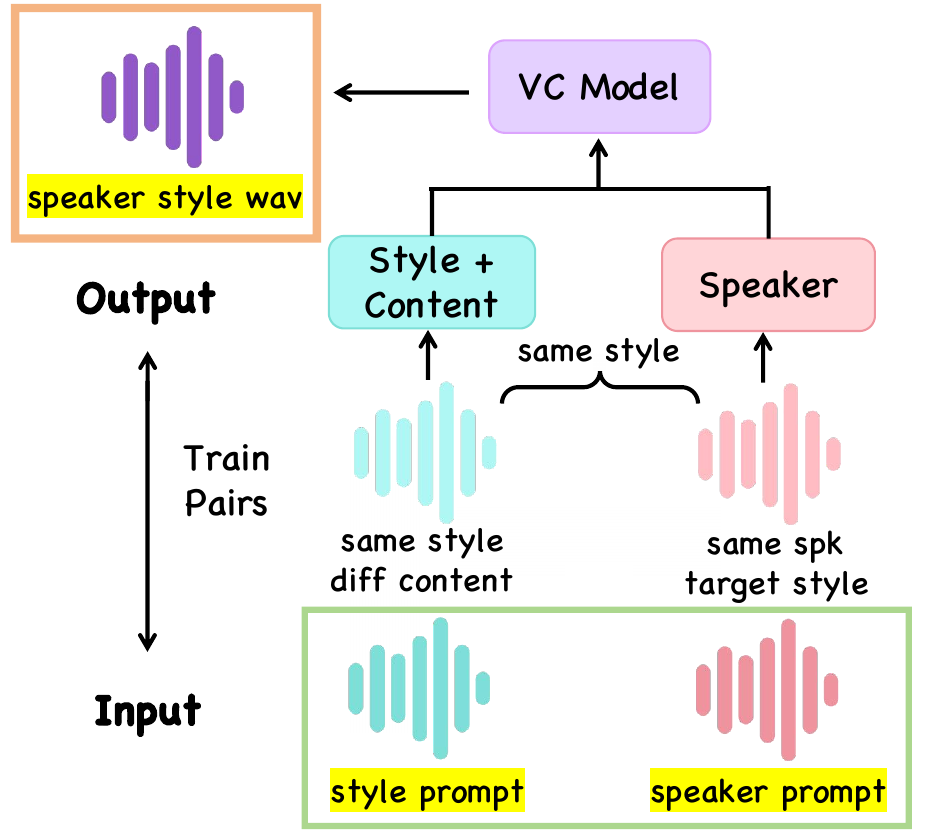}
    \caption{An illustration of the training paradigm.}
    \label{fig:data-driven}
    \vspace{-2mm}
\end{figure}

To enable arbitrary “timbre × style” synthesis, paired ground-truth speech is inherently scarce. We therefore construct pseudo ground-truth via VC and adopt a data-driven training framework to disentangle style and timbre (Fig.~\ref{fig:data-driven}). 

The construction implements three leakage-prevention mechanisms:
\textbf{(a) Avoid timbre leakage:} After selecting the speaker prompt (timbre), reusing the same speaker prompt for the style prompt leaks timbre; we instead select a different speaker for the style prompt.
\textbf{(b) Avoid content leakage:} Since the VC model preserves linguistic content, using the same utterance as both style prompt and VC target causes content leakage. Instead, we choose another utterance from the same speaker with high style similarity be the VC target, preventing textual/content leakage.
\textbf{(c) Avoid style leakage:} To prevent style leakage, where the model acquires style from the speaker prompt, we take speech of the same speaker but a different style as the VC timbre reference.

Table~\ref{tab:compara_real_and_ours} compares retrieved real speech with our VC-constructed pseudo targets. Our construction substantially improves SIM-Style while enabling arbitrary timbre-style combinations. SIM-Spk and SIM-Style are computed against the speaker and style prompts, respectively. We further filter pseudo targets by WER and speaker similarity.


\newcolumntype{Y}{>{\centering\arraybackslash}X} 

\begin{table}[t]
\captionsetup{skip=2pt}
\centering
\caption{Comparison between Same-Speaker Real Ground Truth and Our Constructed Pseudo Ground Truth.}
\label{tab:compara_real_and_ours}
\renewcommand{\arraystretch}{0.85} 
\begin{tabularx}{\columnwidth}{@{}lYY@{}}
\toprule
Dataset          & SIM-Spk & SIM-Style \\
\midrule
Real Speaker GT  & 82.50 & 71.20 \\
Our Pseudo GT     & 77.60 & 92.85 \\
\bottomrule
\end{tabularx}
\vspace{-1em}
\end{table}

\begin{figure*}[t!]
    \centering
    \includegraphics[width=0.67\linewidth]{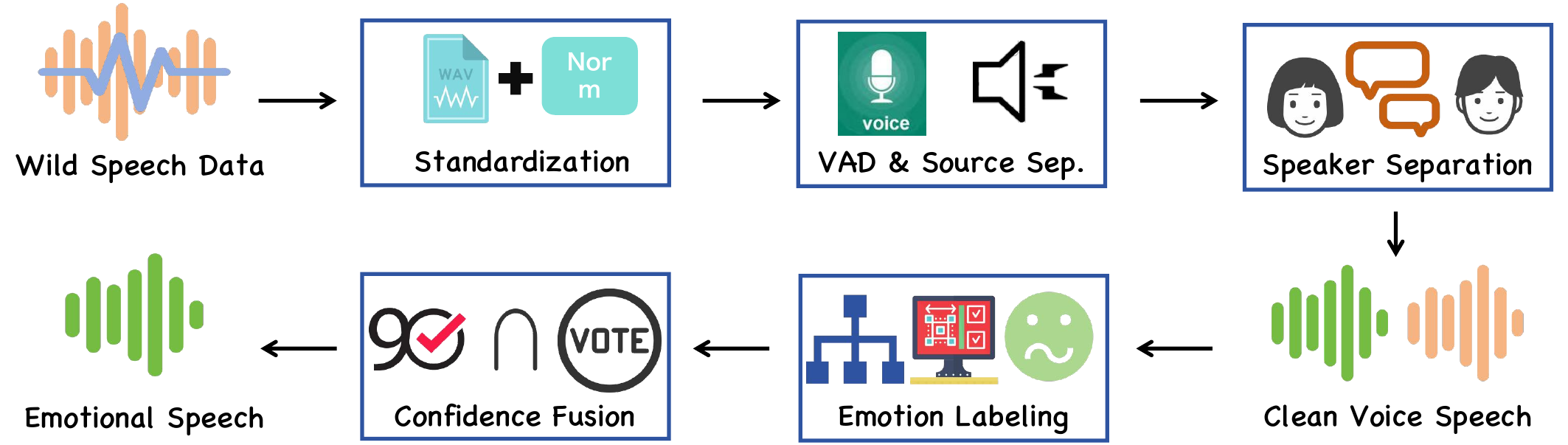}
    \caption{An illustration of the proposed dataset pipeline to create the EmoData dataset.}
    \label{fig:enter-label}
    \vspace{-0.8em}
\end{figure*}

\subsection{EmoData: Expressive Large-Scale Benchmark}
To support expressive and multilingual TTS, we construct EmoData, the first thousand-hour multilingual dataset as a large expressive style space for multilingual emotional speech. It is constructed via a fully automated and scalable pipeline, covering six emotions: happy, angry, sad, surprised, disgusted, and neutral. With large-scale data and diverse expressive variations for each emotion, EmoData enables both coarse-grained (category) and fine-grained (intensity/prosody) control over emotional expression in synthesis. Compared with existing emotional speech resources \cite{DBLP:journals/lre/BussoBLKMKCLN08, adigwe2018emotional, Xin_2024, abilbekov2024kazemottsdatasetkazakhemotional,zhou2022emotionalvoiceconversiontheory}, EmoData offers significant improvements in scale, naturalistic (wild data), and multilingual coverage. It also achieves semantic–prosodic consistency and emotional intensity comparable to studio recordings.

The overall construction process as shown in Figure~\ref{fig:enter-label} adopts a coarse-to-fine cascaded filtering strategy. We begin by applying loudness normalization and resampling to the raw audio. To handle complex background noise in wild scenarios, we employ industrial-grade source separation models to isolate clean speech, followed by a robust speaker diarization system to cluster and filter speakers by duration.
\indent For emotion labeling, we introduce a confidence-aware ensemble mechanism. Instead of relying on a single tagger, we fuse predictions from multiple state-of-the-art SER models \cite{ma2023emotion2vecselfsupervisedpretrainingspeech,geng2025osumadvancingopenspeech,an2024funaudiollmvoiceunderstandinggeneration} with our internal validators. {The recognition models are fine-tuned in multiple languages, improving accuracy and robustness for better cross-lingual performance.} To ensure high expressiveness, we implement a dual-check strategy: samples are retained only if (1) the acoustic emotion confidence exceeds a high threshold, and (2) the linguistic sentiment aligns with the acoustic prosody. The pipeline uses automated filtering tuned by human feedbacks, using sample reviews to determine the best thresholds. This pipeline effectively eliminates semantic-prosodic mismatches while scaling to thousands of hours without manual annotation. 

To ensure compliance and privacy, all speech were collected from publicly available sources and subjected to rigorous automated de-identification to remove personally identifiable information in accordance with ethical research standards.

\begin{table}[htbp]
\centering
\footnotesize
\caption{Comparison of publicly emotional datasets.}
\label{table:dataset_statistics}
\renewcommand{\arraystretch}{1.1} 
\footnotesize 
\resizebox{\columnwidth}{!}{ 
\begin{tabular}{
  @{\hspace{0pt}}l@{\hspace{0.6pt}}  
  c@{\hspace{0.5pt}}                 
  >{\centering\arraybackslash}p{1cm}@{\hspace{0pt}} 
  >{\centering\arraybackslash}p{1cm}@{\hspace{0pt}} 
  c@{\hspace{1pt}}                 
  >{\centering\arraybackslash}p{1cm}@{\hspace{0pt}} 
  >{\centering\arraybackslash}p{1cm}@{\hspace{0pt}} 
  >{\centering\arraybackslash}p{1cm}@{\hspace{0pt}} 
}
\toprule
\textbf{Dataset} & \parbox{1cm}{\centering \textbf{Language} \\ \textbf{Num}} & \parbox{1cm}{\centering \textbf{Num} \\ \textbf{Speakers}} & \parbox{1cm}{\centering \textbf{Dur} \\ \textbf{(hrs)}}  & \parbox{1cm}{\centering \textbf{SR} \\ \textbf{(kHz)}} & \parbox{1cm}{\centering \textbf{Wild/} \\ \textbf{Studio}} \\ 

\midrule
IEMOCAP\cite{DBLP:journals/lre/BussoBLKMKCLN08}             & EN & 10    & 12.6  & \textbf{48} & S \\
EMOV-DB\cite{adigwe2018emotional}              & EN,FR & 5  & 7.2 (E.) & 16 & S \\
ESD\cite{zhou2022emotionalvoiceconversiontheory}             & EN,ZH & 20  & 30 (E.) & 16 & S \\
KazEmo\cite{abilbekov2024kazemottsdatasetkazakhemotional}             & KA & 3  & 74.85 & 44.1 & S \\
JVNV\cite{Xin_2024}             & JA & 4  & 3.94 & 44.1 & S \\
\midrule
EmoData      & \parbox{1.5cm}{\centering \textbf{EN,ZH,JA,} \\ \textbf{KO,GE,FR}} & \parbox{1cm}{\centering \textbf{5681}} & \parbox{1cm}{\centering \textbf{1012}}  & 44.1 & \textbf{W} \\
\bottomrule
\end{tabular}}
{
(E.) indicates value estimated from the paper based on average.}
\vspace{-6mm}
\end{table}

\subsection{Style-RAG: Intent-Based Acoustic Grounding}
While EmoData provides the style space, users still need a practical way to select style prompts. Style-RAG addresses this by grounding natural-language style intent into acoustic references from the EmoData style space. We build a high-quality caption--wav database that pairs aligned text descriptions and style speech. All style utterances are preprocessed and validated for expressiveness, so they can be directly used for inference without extra user effort. The Style-RAG module utilizes the contrastive score \cite{wu2024largescalecontrastivelanguageaudiopretraining} to compute both cross-modal similarity and automatically retrieves the most suitable style speech for user-provided style caption. This delivers accurate, diverse style prompts while reducing manual selection, and enables the system to flexibly adapt synthesis style.

\section{Experiments}

\subsection{Evaluation Metrics}

We evaluate both objective and subjective metrics.
 
\textbf{Objective Metrics:} We report Word Error Rate (WER) for intelligibility, Speaker Similarity (SIM-Spk) \cite{Chen_2022} for timbre consistency against the target prompt. And Style Similarity (SIM-Style) calculates the cosine similarity of style embeddings \cite{ma2023emotion2vecselfsupervisedpretrainingspeech} between the synthesized audio and reference style.

\textbf{Subjective Metrics:} We use Speech-Naturalness MOS (SN-MOS) to rate naturalness and style-semantics alignment. For perceptual similarity, we conduct Speaker-Similarity MOS (SS-MOS) to assess speaker resemblance in single-prompt settings, and Style-Matching MOS (SM-MOS) for timbre-independent style evaluation in dual-prompt settings.

\subsection{Dataset Evaluation}

We evaluate EmoData through both subjective perception tests and objective fine-tuning ablation tests.

\begin{figure}
    \centering
    \includegraphics[width=0.67\linewidth]{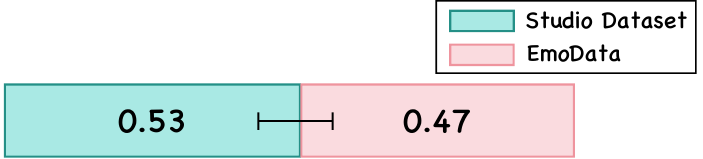}
    \caption{AB Test Between Studio and EmoData}
    \label{fig:placeholder}
\vspace{-4mm}
\end{figure}

\textbf{Subjective Quality and Expressiveness.} First, we verify whether our mined wild data can match the quality of traditional studio recordings. We conducted A/B preference tests on 180 clips (30 per emotion category) with 15 native speakers. Fig~\ref{fig:placeholder} shows that EmoData achieves perceptual quality comparable to studio datasets. Furthermore, the high style matching rates across all categories in Table~\ref{tab:smr} confirm precise alignment between semantics and prosody. These results demonstrate that our automated pipeline successfully extracts expressive, studio-grade speech from wild sources.

\textbf{Objective Utility and Dataset Ablation.} We fine-tune strong TTS baseline \cite{du2024cosyvoicescalablemultilingualzeroshot} to assess the practical value of EmoData. Table~\ref{tab:ft_cosyvoice} reveals a critical insight: fine-tuning on raw data degrades speech intelligibility due to noise and misalignment. In contrast, training with EmoData significantly improves style similarity while preserving high intelligibility and timbre consistency. This validates EmoData as a high-quality, highly expressive resource, demonstrating its substantial practical contribution to advancing TTS performance.

\begin{table}[ht]
\vspace{-1em}
\captionsetup{skip=2pt}
\centering
\caption{Style Matching Rate(\%)}
\label{tab:smr}
\renewcommand{\arraystretch}{0.8} 
\begin{tabular}{cccccc}
\toprule
Neutral & Happy & Angry & Sad & Surprised & Disgusted \\
\midrule
95.20 & 91.75 & 92.30 & 90.45 & 85.10 & 78.65 \\
\bottomrule
\end{tabular}
\vspace{-4mm}
\end{table}

\begin{table}[ht]
\captionsetup{skip=2pt}
\vspace{-1em}
\centering
\caption{Finetuned Results on EmoData.}
\label{tab:ft_cosyvoice}
\renewcommand{\arraystretch}{0.8} 
\begin{tabularx}{\columnwidth}{@{} 
  >{\hsize=1.45\hsize\raggedright\arraybackslash}X 
  >{\hsize=0.75\hsize}Y 
  >{\hsize=0.9\hsize}Y 
  >{\hsize=0.9\hsize}Y @{}}
\toprule
Dataset          & WER(\%)$\downarrow$ & SIM-Spk$\uparrow$ & SIM-Style$\uparrow$ \\
\midrule
CosyVoice \cite{du2024cosyvoicescalablemultilingualzeroshot}  & 3.05 & 70.68 & 72.51\\
Finetuned w/ Raw     & 3.21 & 71.33  & 71.21\\
Finetuned w/ Ours    & \textbf{2.98} & \textbf{73.04}  & \textbf{78.63}\\
\bottomrule
\end{tabularx}
\vspace{-5mm}
\end{table}

\subsection{Style-Timbre Disentangled Synthesis}

We evaluate our framework on the LibriTTS test \cite{zen2019librittscorpusderivedlibrispeech} by randomly sampling 300 speaker prompts and style prompts. The evaluation is conducted across two distinct scenarios: standard single-prompt zero-shot synthesis and dual-prompt disentangled synthesis with ablations.

\textbf{Zero-Shot Synthesis.} Table~\ref{tab:sota_compare_base} presents the results where timbre and style are conditioned on the same speaker prompt, even for models supporting dual inputs. The columns Wav P. and Text P. denote style controllability via speech prompts and text prompts, respectively. Our pre-trained baseline demonstrates superior performance compared to SOTA models, where E3-System also demonstrates robust foundational TTS capabilities. It performs on par with strong state-of-the-art baselines \cite{du2024cosyvoicescalablemultilingualzeroshot, du2024cosyvoice2scalablestreaming, guo2025fireredttsfoundationtexttospeechframework, wang2024maskgctzeroshottexttospeechmasked, wang2025sparkttsefficientllmbasedtexttospeech} in terms of intelligibility and overall naturalness, while slightly outperforming them in speaker similarity. Note that RAG is designed for retrieval-based style conditioning and is therefore not evaluated in the single-prompt setting. {Ablation results shows a downsampling factor of 2 yields a ~1.8x speedup and slight quality gains over the single-AR baseline, whereas larger factors degrade performance.}

\textbf{Disentangled Synthesis and Ablation.} Table~\ref{tab:ablation_style_control} evaluates dual-prompt disentangled synthesis, enabling arbitrary timbre $\times$ style combinations with distinct speaker and style prompts. This scenario highlights a common trade-off: intense emotion transfer often degrades speaker identity. While competitive dual-prompt baselines \cite{zhang2025vevocontrollablezeroshotvoice,li2026discospeechcontrollablezeroshotspeech} achieve strong style matching, they often suffer from source timbre leakage. In contrast, E3-System enforces stricter disentanglement through our disentangled codec and data-driven strategy, achieving an superior balance between style alignment and speaker preservation, outperforming SOTA baselines in both speaker and style similarity.

Ablation studies further validate our designs. Replacing our disentangled strategy with a standard adversarial approach causes a sharp decline in style similarity, proving the necessity of our strict leakage-prevention construction. 
Finally, comparing Ours w/ RAG against the oracle setting (Ours w/o RAG) reveals that while direct speech prompts yield higher stylistic accuracy, our RAG module achieves comparable style similarity and speech quality. This proves Style-RAG offers intuitive, manual-free text control without sacrificing synthesis quality.

\newcolumntype{Y}{>{\centering\arraybackslash}X}
{

}

{
}

\vspace{-1em}
\begin{table}[ht]
\centering
\caption{Evaluation of zero-shot synthesis and style controllability in single-prompt setting. NA indicates \textit{“Not applicable”} in a single-prompt setting.}
\vspace{-1em}
\label{tab:sota_compare_base}
\hspace*{-0.015\columnwidth}\resizebox{1.03\columnwidth}{!}{%
  \setlength{\tabcolsep}{4pt} 
  \renewcommand{\arraystretch}{1} 
  \begin{tabular}{@{} l c c c c c c @{}} 
    \toprule
    \textbf{Model} & \textbf{Wav P.} & \textbf{Txt P.} & \textbf{WER$\downarrow$} & \textbf{SIM$\uparrow$} & \textbf{SS-MOS$\uparrow$} & \textbf{SN-MOS$\uparrow$} \\
    \midrule
    
    \multicolumn{7}{c}{\textit{Text-Prompted TTS (w/o Zero-Shot Voice Cloning)}} \\
    \midrule
    PromptTTS \cite{guo2022promptttscontrollabletexttospeechtext} & \ding{55} & \checkmark & 4.25 & NA & NA & 3.51$\pm$0.12 \\
    InstructTTS \cite{yang2023instructttsmodellingexpressivetts} & \ding{55} & \checkmark & 6.78 & NA & NA & 3.47$\pm$0.13  \\
    \midrule
    
    \multicolumn{7}{c}{\textit{Zero-Shot TTS (w/o Disentangled Style)}} \\
    \midrule
    FireRedTTS \cite{guo2025fireredttsfoundationtexttospeechframework} & \ding{55} & \ding{55} & 3.79 & 60.02 & 3.51$\pm$0.09 & 3.72$\pm$0.10 \\
    MaskGCT \cite{wang2024maskgctzeroshottexttospeechmasked} & \ding{55} & \ding{55} & 2.63 & 69.70 & 3.75$\pm$0.16 & 3.74$\pm$0.12 \\
    SparkTTS \cite{wang2025sparkttsefficientllmbasedtexttospeech} & \ding{55} & \ding{55} & 2.56 & 65.54 & 3.66$\pm$0.11 & 3.78$\pm$0.14 \\
    CosyVoice \cite{du2024cosyvoicescalablemultilingualzeroshot} & \ding{55} & \ding{55} & 2.96 & 71.30 & 3.79$\pm$0.13 & 3.70$\pm$0.12 \\
    CosyVoice2\cite{du2024cosyvoice2scalablestreaming} & \ding{55} & \ding{55} & 2.48 & 74.65 & \textbf{3.89$\pm$0.13} & 3.95$\pm$0.13 \\
    \midrule
    
    Our Pretrained & \ding{55} & \ding{55} & 2.41 & 73.88 & 3.81$\pm$0.15 & 3.87$\pm$0.13 \\
    \midrule
    
    \multicolumn{7}{c}{\textit{Explicitly Controllable TTS (Zero-Shot Voice Cloning w/ Disentangled Style)}} \\
    \midrule
    Ours w/o RAG & \checkmark & \ding{55} & \textbf{2.33} & \textbf{74.82} & 3.87$\pm$0.14 & \textbf{3.96$\pm$0.11} \\
    Ours w/ RAG & \checkmark & \checkmark &
\multicolumn{4}{c}{NA in single-prompt setting} \\
    \bottomrule
  \end{tabular}%
}
\end{table}

\begin{table}[ht]
\centering
\small
\setlength{\tabcolsep}{3pt}      
\renewcommand{\arraystretch}{1.05}
\vspace{-2em}
\captionsetup{skip=2pt}
\caption{Evaluation of explicit style control and disentanglement in dual-prompt setting.}
\label{tab:ablation_style_control}
\begin{tabular}{@{}
  p{0.26\columnwidth}  
  >{\centering\arraybackslash}p{0.10\columnwidth}  
  >{\centering\arraybackslash}p{0.10\columnwidth}  
  >{\centering\arraybackslash}p{0.10\columnwidth}  
  >{\centering\arraybackslash}p{0.16\columnwidth}  
  >{\centering\arraybackslash}p{0.16\columnwidth}@{} 
}
\toprule
\textbf{Model} & \textbf{WER$\downarrow$}& \textbf{SIM-} \newline \textbf{Spk$\uparrow$} & \textbf{SIM-} \newline \textbf{Style$\uparrow$} & \textbf{SM-MOS$\uparrow$} & \textbf{SN-MOS$\uparrow$} \\
\midrule
Vevo-TTS \cite{zhang2025vevocontrollablezeroshotvoice} & 3.52 & 65.16 & 85.06 & 3.90$\pm$0.15 & 3.87$\pm$0.21 \\
IndexTTS2 \cite{zhou2025indextts2breakthroughemotionallyexpressive} & 2.63 & 69.85 & 86.73 & 3.96$\pm$0.13 & 3.88$\pm$0.13  \\
DisCo \cite{li2026discospeechcontrollablezeroshotspeech} & 3.16 & 63.68 & 87.25 & 4.00$\pm$0.11 & 3.91$\pm$0.09  \\
\midrule
Ours (GRL) & 3.08 & 59.02 & 72.50 & 3.61$\pm$0.13 & 3.74$\pm$0.12 \\
Ours w/o RAG & \textbf{2.51} & \textbf{70.15} & \textbf{89.22} & \textbf{4.08$\pm$0.14} & \textbf{3.98$\pm$0.11} \\
Ours w/ RAG & 2.93 & 69.08 & 86.33 & 3.97$\pm$0.12 & 3.96$\pm$0.10 \\
\bottomrule
\end{tabular}
\end{table}\vspace{-2\baselineskip}

\section{Conclusion}
This paper presents E3-System, a unified framework for expressive zero-shot TTS that integrates large-scale expressive data, intent-based acoustic grounding, and timbre-style disentangled synthesis. Experiments demonstrate competitive intelligibility and speaker preservation while achieving strong style consistency and naturalness. Future work will extend E3-System to broader multimodal control.
\vfill\pagebreak

\bibliographystyle{IEEEtran}
\begingroup
\renewcommand{\baselinestretch}{0.76}
\small

\makeatletter
\patchcmd{\thebibliography}
  {\section{\refname}}
  {\section*{\refname}}
  {}
  {}
\makeatother

\bibliography{refs}
\endgroup
\end{document}